# Enabling Data Discovery through Virtual Internet Repositories

Giriprakash Palanisamy[1], Ranjeet Devarakonda[1], Bruce Wilson[1], Jim Green[2]

(1) Environmental Sciences Division, Oak Ridge National laboratory, Oak Ridge, TN
(2) On subcontract from Information International Associates, Oak Ridge, TN

## Abstract

Mercury is a federated metadata harvesting, search and retrieval tool based on both open source and software developed at Oak Ridge National Laboratory. It was originally developed for NASA, and the Mercury development consortium now includes funding from NASA, USGS, and DOE. A major new version of Mercury was developed during 2007. This new version provides orders of magnitude improvements in search speed, support for additional metadata formats, integration with Google Maps for spatial queries, support for RSS delivery of search results, among other features. Mercury provides a single portal to information contained in disparate data management systems. It collects metadata and key data from contributing project servers distributed around the world and builds a centralized index. The Mercury search interfaces then allow the users to perform simple, fielded, spatial and temporal searches across these metadata sources. This centralized repository of metadata with distributed data sources provides extremely fast search results to the user, while allowing data providers to advertise the availability of their data and maintain complete control and ownership of that data.

## Introduction

As the number of scientific datasets created by various research projects dramatically increases, existing site specific data discovery systems are less effective and these datasets are found in thousands of repositories located around the world which makes it tedious for the user to search. Virtual observatories and distributed metadata search and data discovery systems are helping the scientists search those repositories to find and access the required data (Todd 2008). Distributed/virtual metadata systems typically harvest these metadata from various data providers and make it available through a single search system. Distributed Active Archive Center [REF: ORNL DAAC], at Oak Ridge National Laboratory (ORNL DAAC) developed a distributed metadata harvesting, search and data discovery system called Mercury [REF: Mercury], which was originally developed for NASA to search biogeochemical data that are archived at the ORNL DAAC center. Mercury system provides a single portal to information contained in disparate data management systems. It provides free text, fielded, spatial, temporal and keyword browse tree search capabilities. Mercury allows individuals and database managers to distribute their data while maintaining complete control and ownership. Mercury has been used by various scientific projects that are funded by NASA, USGS and DOE (ORNL DAAC, NBII, DADDI, LBA, LTER, NARSTO, CDIAC, OCEAN, I3N, IAI, ARM). Mercury is operated as a consortium and the development and operating costs are shared by these projects. In this paper we discuss Mercury's harvesting models, indexing techniques, and various search services that are available through the Mercury system.

## Methods and Techniques

Mercury supports various metadata standards including XML, Z39.50, FGDC, Dublin-Core, Darwin-Core, EML, and ISO-19115. The new Mercury system is based on open source and Service Oriented Architecture and provides multiple search services.

Mercury architecture includes different components, a harvester, an indexing tool, and a user interface. Mercury's harvester operates in two different models, 1) virtual internet database and 2) virtual aggregate database. The virtual internet database model organizes a new collection of data from informal systems spread across the internet, in this, typically the data providers or the principal investigators create the metadata for their datasets and place these metadata in a publically accessible place such as a web directory or FTP directory. Mercury then harvests these metadata and builds a centralized index and makes it available for the Mercury search user interface.

In the virtual aggregate database model, Mercury harvests information from existing formal disparate database management systems (DBMS). In this, the metadata exists in remote databases, custom export programs can be easily written to extract the metadata from these DBMS and the metadata are saved in xml files. Mercury then harvests the extracted metadata files and builds a centralized index for metadata searching (Figure 1). Some Mercury instances are using both these models to harvest the metadata. Mercury development team is currently working on enabling a metadata harvesting service using the Open Archives Initiative (OAI).

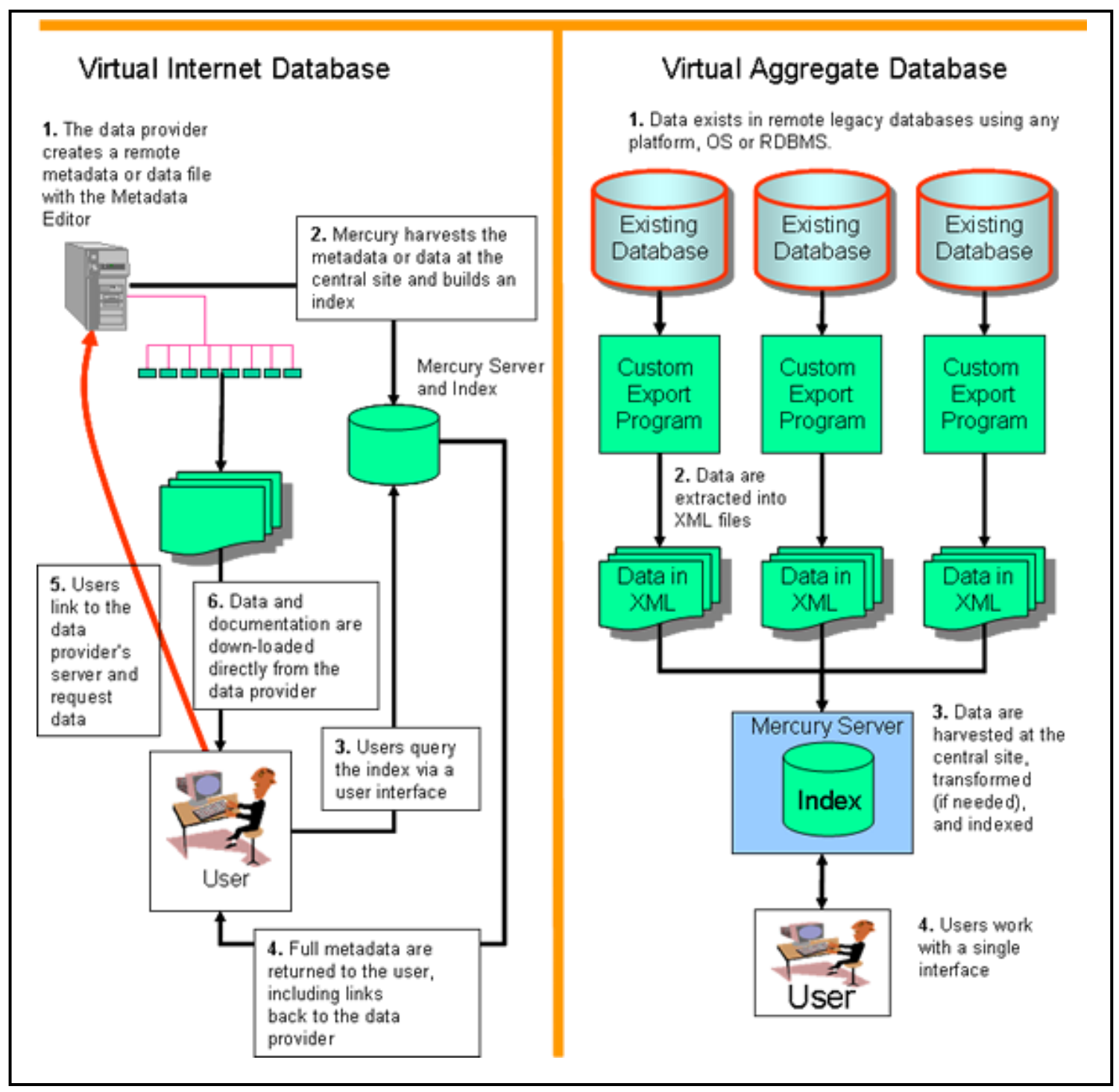

Figure 1. Mercury metadata harvesting architecture

The new Mercury, which was recently re-designed using various open source tools, encompasses changing the indexing and searching interface, from a proprietary implementation to an open source search server. The open source Apache project -codename Lucene [REF: Lucene] which is a free/open source information retrieval library, is used in conjunction with Solr [REF: Solr] which is an open source enterprise search server based on the Lucene. Solr is another Apache open source project that extends the functionality of Lucene, given proper consideration to numeric types, dynamic fields, unique keys, and faceted searching. An example to clarify what this means: Solr gives the developer the ability to give special treatment to specific geotemporal coordinates. Special information that is used in an advanced search can be treated properly using

Solr, as opposed to be buried among the competing rankings given by the Lucene to all the metadata content.

## Results and Discussion

Typical Mercury user interface provides three different search capabilities. 1) simple search, 2) advanced search and 3) Web browse tree search. In the simple search option, users can perform a full text search. In the advanced search option, users will be able to search by specifying keywords, time period, spatial extend and the data provider information. Figure 2 is a snapshot of the Mercury advanced search interface used in ORNL DAAC [REF ORNL Mercury].

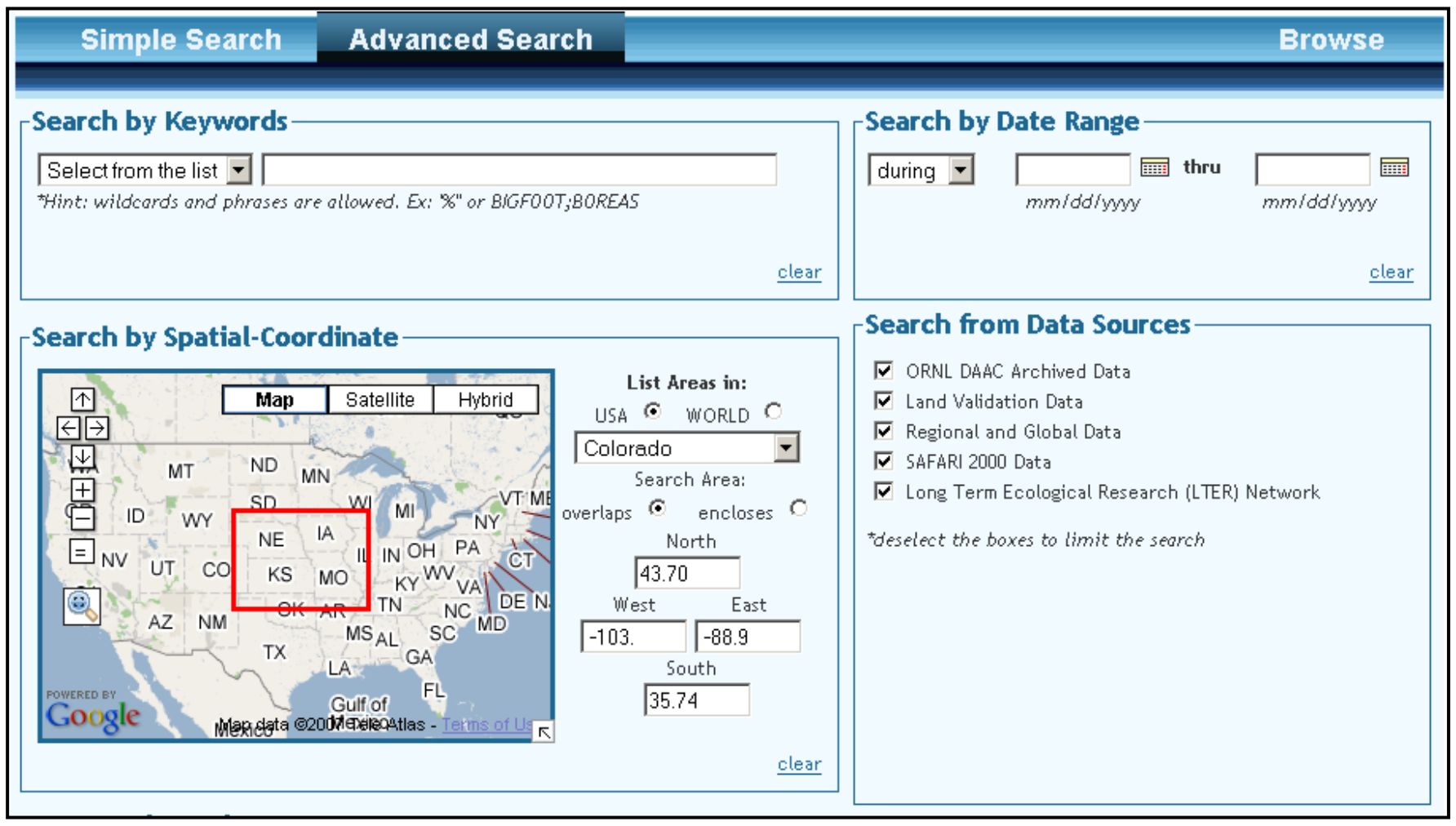

Figure 2. A snapshot of the ORNL-DAAC advance search interface

In the web browse tree search option, users will be able to drill down to their metadata of interest using a hierarchical keyword tree (figure 3).

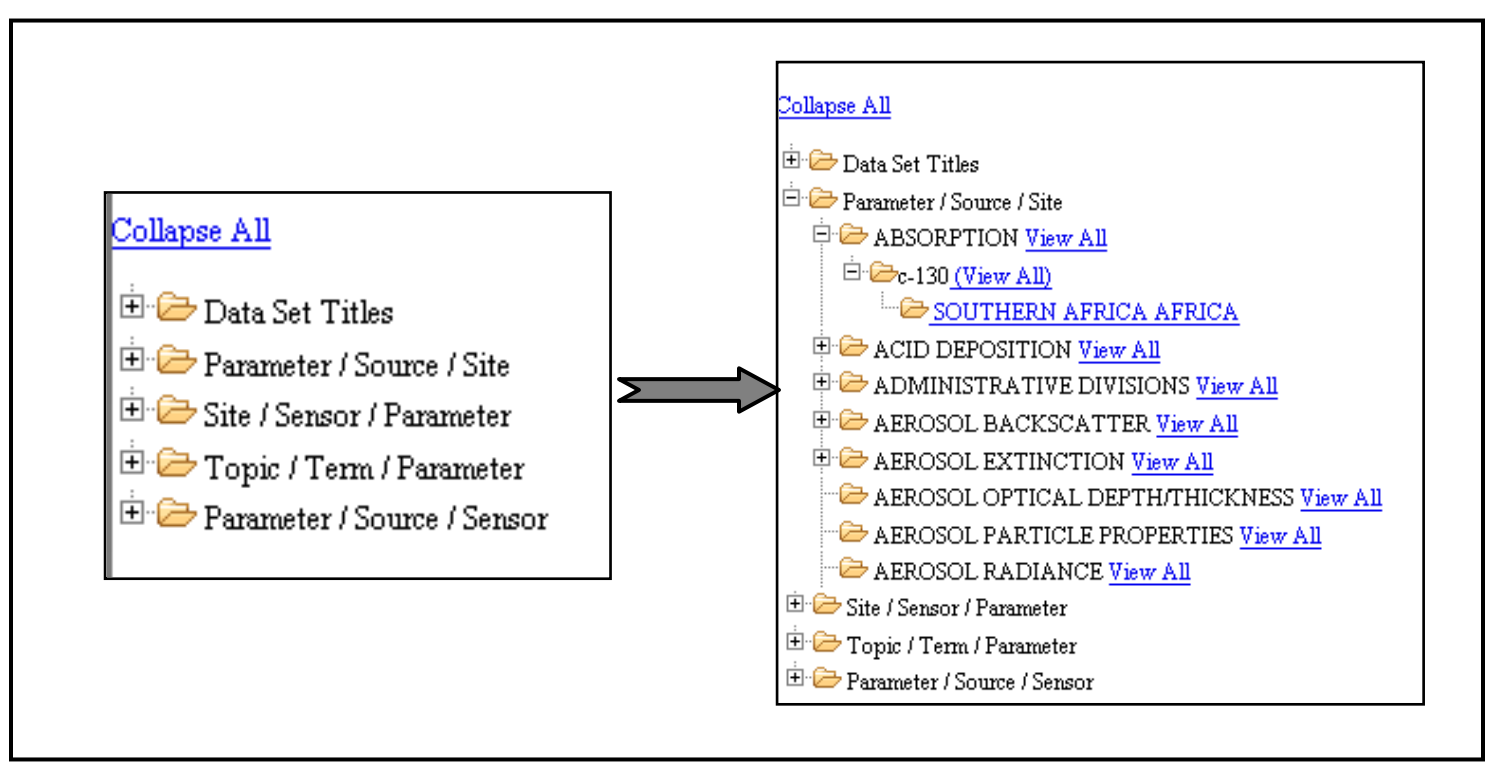

Figure 3. A snapshot of the ORNL-DAAC browse tree search

Once the users enter their search criteria and perform the search, the results summary page displays the total number of records found for the search and option for filtering the search results using logical groupings (by data providers, parameter, sensor, topic, project etc.). The summary page also allows the users to sort the results based on the search relevancy, period of record, source and project. The page shows push buttons in the top right to create an RSS feed, a bookmark or an email for these results. RSS or bookmarks enable refreshing the query matches periodically without the hassle of recreating the query.

The bottom of the summary page shows the results, snippets of the records that match the search/browse criteria, and a link to the full metadata and a link to access the associated data. The stars shown at the bottom of each record indicate the relative relevance of the matched criteria. The snippet includes the title and study date range, source provenance and excerpts from the abstract (Figure 4).

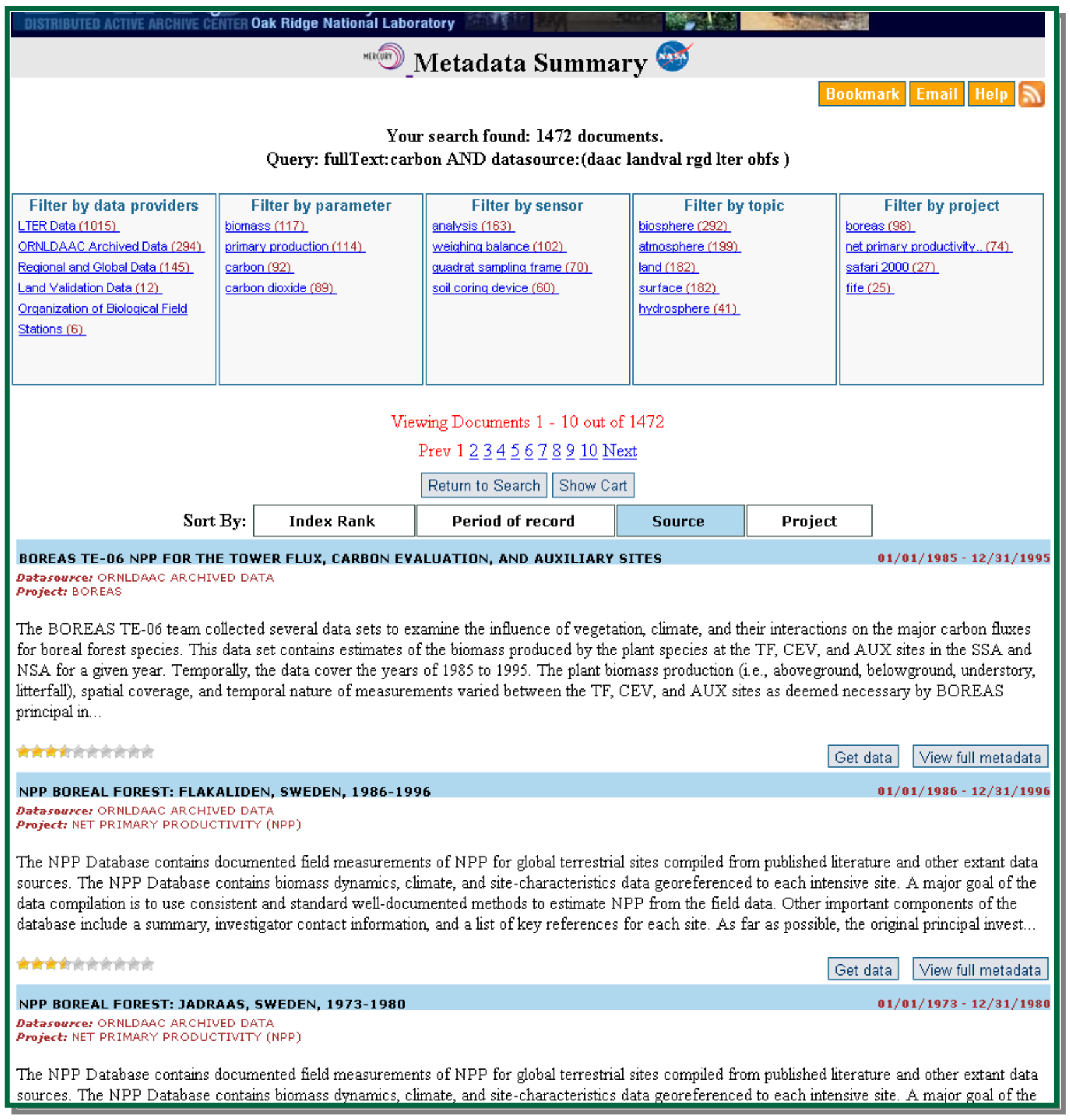

Figure 4. A typical look at the query results page

When the user clicks the "View Full Metadata" link found on the summary page, the Mercury metadata report's page will be displayed. This page offers two styles to display a full metadata record. The Mercury by default offers a classic, well organized redux style at the full records page and additionally, it offers what it is known as the FGDC style, which would be very familiar to those who use the ESRI tools or that have used the previous mercury. It is plain text divided in 6 sections, with the underlying hierarchy preserved as indentation.

Mercury also provides the harvested metadata to other applications (e.g., Google, NASA Global Change Master Directory, NBII Biobot).The National Biological Information Infrastructure [REF: NBII] Clearinghouse [REF: NBII CH] consumes the search results as portlets in their NBII portal web application, which is another way of displaying the customized search results in external web pages. Global Forestry Information Services [REF: GFIS] which is partnering with NBII Clearinghouse is harvesting all the forest related metadata records as RSS service and exposing those records through their search system.

**Conclusions:**

Mercury serves more than 50,000 metadata records through its various project specific user interfaces. Mercury supports various metadata standards including XML, Z39.50, FGDC, Dublin-Core, Darwin-Core, EML, and ISO-19115. The new Mercury system is based on open source and Service Oriented Architecture and provides multiple search services including; user interface search tools, RSS services for search results, bookmark search results, portlets supports.

**Acknowledgements:**

Mercury consortium is funded by NASA, USGS DOE for a consortium of projects NASA, USGS and DOE including ORNL DAAC, NBII, DADDI, LBA, LTER, NARSTO, CDIAC, OCEAN, I3N, and IAI.